\documentclass[amssymb,amsmath,aps,showpacs,floatfix,nofootinbib,showpacs,12pt,prd]{revtex4}

\usepackage{graphicx}
\usepackage{latexsym}
\usepackage{color}
\def\be{\begin{equation}}
\def\ee{\end{equation}}
\def\bea{\begin{eqnarray}}
\def\eea{\end{eqnarray}}

\begin{document}

\title{Neutrino emission from dark matter annihilation/decay in light
of cosmic $e^{\pm}$ and $\bar{p}$ data}

\author{Jie Liu${}^{a}$}
\author{Qiang Yuan${}^{b}$}
\author{Xiaojun Bi${}^{b}$}
\author{Hong Li${}^{a,c}$}
\author{Xinmin Zhang${}^{a,c}$}

\affiliation{${}^a$Theoretical Physics Division, Institute of High
Energy Physics, Chinese Academy of Science, P.O.Box 918-4, Beijing
100049, P.R.China}

\affiliation{${}^b$Key Laboratory of Particle Astrophysics,
Institute of High Energy Physics, Chinese Academy of Science,
P.O.Box 918-3, Beijing 100049, P.R.China}

\affiliation{${}^c$Theoretical Physics Center for Science Facilities
(TPCSF), Chinese Academy of Science, P.R.China}


\begin{abstract}

A self-consistent global fitting method based on the Markov Chain Monte
Carlo technique to study the dark matter (DM) property associated with the
cosmic ray electron/positron excesses was developed in our previous work.
In this work we further improve the previous study to include the
hadronic branching ratio of DM annihilation/decay. The PAMELA 
$\bar{p}/p$ data are employed to constrain the hadronic branching ratio.
We find that the $95\%$ ($2\sigma$) upper limits of the quark branching 
ratio allowed by the PAMELA $\bar{p}/p$ data is $\sim 0.032$ for DM
annihilation and $\sim 0.044$ for DM decay respectively. This result shows 
that the DM coupling to pure leptons is indeed favored by the current
data. Based on the global fitting results, we further study the neutrino 
emission from DM in the Galactic center. Our predicted neutrino flux is 
some smaller than previous works since the constraint from $\gamma$-rays
is involved. However, it is still capable to be detected by the forth-coming
neutrino detector such as IceCube. The improved points of the 
present study compared with previous works include: 1) the DM 
parameters, both the particle physical ones and astrophysical ones, 
are derived in a global fitting way, 2) constraints from various species 
of data sets, including $\gamma$-rays and antiprotons are included, 
and 3) the expectation of neutrino emission is fully self-consistent.

\end{abstract}


\maketitle


\section{Introduction}
\label{Int}

In a previous work (\cite{Liu:2009sq}, Paper I) we have developed a Markov 
Chain Monte Carlo (MCMC) code to fit the parameters of dark matter 
(DM) models, which are proposed to explain the recent reported abnormal 
excesses of cosmic ray (CR) positrons and electrons by PAMELA 
\cite{Adriani:2008zr}, ATIC \cite{Chang:2008zzr}, 
HESS \cite{Aharonian:2008aa,Aharonian:2009ah}
and Fermi-LAT \cite{Abdo:2009zk}. One assumption adopted in that work 
is that the DM particles only couple with leptons. To extend the 
discussion including hadronic channels will be natural and necessary.
Actually the non-excess of PAMELA $\bar{p}/p$ data \cite{Adriani:2008zq} 
have been studied in many works to set constraints on the hadronic
couplings of DM particles (e.g., \cite{Cirelli:2008pk,Yin:2008bs,
Donato:2008jk}). However, these studies can only give some illustration
constraints on the model parameters instead of a full scan of the possible
parameter space. Our MCMC fitting scheme makes it possible to fully scan
the high-dimensional parameter space and derive the model-independent
constraints from the data directly. 

Shortly after the proposal of using DM annihilation to account for the
data, the accompanied $\gamma$-ray and radio emission are investigated
as constraint and/or future probe of the current models (e.g., 
\cite{Mack:2008wu,Bertone:2008xr,Zhang:2008tb,Bergstrom:2008ag}). Besides the 
photon emission, neutrino can also be regarded as an another 
probe to test the DM interpretation of the data \cite{Beacom:2006tt,
Yuksel:2007ac,Liu:2008ci,Spolyar:2009kx,Buckley:2009kw}. 
In Ref. \cite{Liu:2008ci} it is shown 
that Antares and IceCube are promising to detect the neutrino signals 
of DM annihilation in the Galactic center (GC) or the massive subhalo 
for models to explain PAMELA and ATIC data, assuming an NFW profile 
\cite{Navarro:1996gj} of DM spatial distribution. It is also claimed 
that for annihilation DM scenario the IceCube/DeepCore detector can 
explore much of the parameter space to explain the PAMELA and 
Fermi-LAT $e^{\pm}$ data, still for an NFW profile \cite{Spolyar:2009kx}. 
For decaying DM scenario IceCube/DeepCore also has 
the potential to exclude some of the parameter space, depending on 
the final states and DM profile \cite{Buckley:2009kw}. In all of these
works, the expectation of neutrino signals will strongly depend on the 
annihilation/decay final states and DM profile, and the constraints
from various kinds of data are not taken into account simultaneously. 

Based on the above points, in this work we improve our MCMC
code to include the hadronic branching ratio, and then we  
re-examine the neutrino signals according to the parameter sets
derived in MCMC calculations. We include the positron fraction
data from PAMELA \cite{Adriani:2008zr}, the electron + positron 
data from Fermi-LAT and HESS \cite{Abdo:2009zk,
Aharonian:2008aa,Aharonian:2009ah}, the diffuse $\gamma$-ray data
of the GC ridge from HESS \cite{Aharonian:2006au}, and the 
antiproton-proton ratio data from PAMELA \cite{Adriani:2008zq} in
this MCMC study.

The paper is organized as follows. In Sec. II we introduce the
production and propagation of the CR electrons/positrons and
antiprotons. In Sec. III we give the MCMC fitting results of DM
model parameters, for both annihilation and decaying scenarios.
The neutrino signals from GC are discussed in Sec. IV. Finally, 
Sec. V is the summary.

\section{Production and propagation of $e^{\pm}$ and $\bar{p}$}

\subsection{Solution of propagation equation}

In the Galaxy the transport of charged particles is affected by
several processes. The scattering off random magnetic fields will
lead to spatial and energy diffusions. The stellar wind may also
blow away the CRs from the Galactic plane. In
addition, interactions of CR particles with the interstellar
radiation field (ISRF) and/or the interstellar medium (ISM) can
result in continuous and catastrophic energy losses. Since the
detailed processes affect the propagation are species-dependent,
the treatments for positrons and antiprotons are separated. For
the transport processes we take a spatial independent diffusion
coefficient $D(E)=\beta D_0{\cal R}^{\delta}$ (where ${\cal
R}=pc/Ze$ is the rigidity) and a constant wind $V_{\rm c}$
directed outwards along $z$. CRs are confined within a cylinder
halo $L$, i.e. the differential density is bound by 
$n(z=\pm L, R_{\rm max})=0$ with $R_{\rm max}$ the scale of
the visible Galaxy. The free parameters of the model are the
halo size $L$ of the Galaxy, the normalization of the diffusion
coefficient $D_0$ and its slope $\delta$, and the constant
galactic wind $V_{\rm c}$.

The propagation equation of CRs can be generally written as
\begin{equation}
-D\Delta N+V_{\rm c}\frac{\partial N}{\partial z}+2h\Gamma_{\rm tot}
\delta(z)N 
+ \frac{\partial}{\partial E}\left(\frac{{\rm d}E}{{\rm d}t}N\right) 
= q({\bf x},E),
\label{prop}
\end{equation}
where $\Gamma_{\rm tot}=\sum_{i=H,He}n_i\,\sigma_i\,v$ is the 
destruction rate of CRs through interaction with ISM in the thin 
gas disk with half height $h\approx 0.1$ kpc, ${\rm d}E/{\rm d}t$ 
is the energy loss rate, and $q({\bf x},E)$ is the source function.
The solutions for electrons/positrons and antiprotons are presented
in the Appendix. More details about the propagation processes and 
the solutions of propagation equation can be found in Refs.
\cite{Maurin:2001sj,Maurin:2006hy,Maurin:2006ps,Lavalle:2006vb,
Lavalle:1900wn}.

For the propagation parameters, we use the medium (referred as ``MED'') 
set of parameters which is derived through fitting the observational 
B/C data given in Ref. \cite{Donato:2003xg}, i.e., $D_0=0.0112$ 
kpc$^{2}$~Myr$^{-1}$, $\delta=0.70$, $V_{\rm c}$=$12$ km s$^{-1}$ 
and the height of the diffusive halo $L=4$ kpc. Note that it has been
pointed out that the ``MED'' setting of parameters is a bit out of the
most recent data \cite{Simet:2009ne,DiBernardo:2009ku}. However, because
the propagation parameters are generally more sensitive to the 
secondary-primary ratio (e.g., B/C and sub-Fe/Fe) and radiactive-stable
isotope ratio (e.g., $^{10}$Be/$^9$Be and $^{26}$Al/$^{27}$Al), instead 
of the electron and positron data which are most concerned here, we will 
take the ``MED'' parameters as benchmark model in this study. The other 
two typical settings of parameters, ``MIN'' and ``MAX'', are discussed 
as systematic uncertainties. For the determination of propagation 
parameters using the MCMC method one can refer to Refs. 
\cite{Putze:2008ps,Putze:2010zn, Trotta:2010mx}.

\subsection{Background}

For all the CRs we consider here, there are backgrounds originated
from the traditional astrophysical sources and/or interactions in
the Milky Way (MW) or the Earth atmosphere\cite{Belotsky:2004st}.

The CR proton and Helium spectra are well measured at the Earth. We
adopt the parameterizations in Ref. \cite{Donato:2008jk} as the
interstellar proton and Helium spectra. Then we calculate the
positrons (together with a secondary electron component with almost
the same flux) and antiprotons produced through interactions between
CR protons, Helium and the ISM. The interaction is restricted in 
a thin disk with half height $\sim 0.1$ kpc and the average ISM 
density is adopted as $\sim 1$ cm$^{-3}$. These parameter choices
were shown to be able to give best-fit to the B/C data 
\cite{Donato:2001ms}. The shapes of secondary positrons and 
antiprotons are fixed to the calculated results, and we further 
employ two normalization parameters $c_{e^+}$ and $c_{\bar{p}}$ 
to describe the uncertainties about the inelastic hadronic cross 
section and propagation effect. $c_{\bar{p}}$ is found to be within 
$1\pm 0.25$ \cite{Donato:2008jk}. For $c_{e^+}$ we restrict it
in a larger range of $1\pm 0.5$ since the positrons are more
sensitively dependent with the propagation \cite{Delahaye:2008ua}.

The background of primary electrons is different from that of positrons 
and antiprotons. Since we do not have exact knowledge of the primary
electron injection spectrum from the acceleration source, we adopt
a 2-parameter power-law function $q_{e^-}=a_{e^-}E_{e^-}^{-b_{e^-}}$
to describe the injection source of primary electrons. This is similar to
the case of GC $\gamma$-rays, which are parameterized by 
$\phi_{\gamma}^{\rm bkg}=a_{\gamma}E_{\gamma}^{-b_{\gamma}}$. $a_{e^-}$,
$b_{e^-}$, $a_{\gamma}$ and $b_{\gamma}$ are fitted in the MCMC
procedure.

There should be two kinds of backgrounds of neutrinos: the atmospheric
background and the astrophysical one. For energies smaller $\sim 100$
TeV the atmospheric background is dominant \cite{Evoli:2007iy}.
Thus we only consider the comparison with atmospheric background in 
this work. The result of atmospheric neutrino background is adopted 
as the direction-dependent calculation \cite{Honda:2006qj} based on 
the muon data.

\subsection{DM contribution}

The source function (emissivity) of DM annihilation or decay to standard
model particles can be written as
\begin{equation}
q^j({\bf r},E)=\sum_{i}B_i\frac{\langle\sigma v\rangle}{2m_{\chi}^2}
\left.\frac{{\rm d}N}{{\rm d}E}\right|_i^j\rho^2({\bf r}),
\label{annisource}
\end{equation}
for DM annihilation, or
\begin{equation}
q^j({\bf r},E)=\sum_{i}B_i\frac{1}{m_{\chi}\tau}\left.\frac{{\rm d}N}
{{\rm d}E}\right|_i^j\rho({\bf r}),
\label{decaysource}
\end{equation}
for DM decay, where $m_\chi$ is the mass of DM particle, $\langle\sigma
v\rangle$ or $\tau$ is the annihilation cross section or decay age of
DM respectively, $B_i$ is the branching ratio to final state channel
$i$, $\left.\frac{{\rm d}N}{{\rm d}E}\right|_i^j$ is the yield spectrum 
of $j$ species for one annihilation or decay for channel $i$, and 
$\rho({\bf r})$ is DM spatial density in the MW halo. In this work we
consider three channels to lepton pairs $e^+e^-,\,\mu^+\mu^-,\,
\tau^+\tau^-$ as well as a quark channel $q\bar{q}$. Since the antiproton
production spectra from decay of various quark flavors do not differ much
from each other, we do not distinguish quark flavors but use an average
result from all the flavors. The spectra $\left.\frac{{\rm d}N}
{{\rm d}E}\right|_i^j$ of positrons, antiprotons, $\gamma$-photons and 
neutrinos from decay of the final state particles of various channels 
are calculated using PYTHIA package \cite{Sjostrand:2006za}.

Similar as in Paper I, we take the density profile of 
the MW halo as the form
\begin{equation}
\rho(r)=\frac{\rho_{s}}{(r/r_s)^{\gamma}(1+r/r_s)^{3-\gamma}},
\label{profile}
\end{equation}
where $\gamma$ represents the central cusp slope of the density
profile, $r_s$ and $\rho_s$ are scale radius and density
respectively. For the MW DM halo, we adopt the total mass to be
$M_{\rm MW} \approx 10^{12}$ M$_{\odot}$ \cite{Xue:2008se} and the
concentration parameter to be $c_{\rm MW}\approx 13.5$
\cite{Bullock:1999he}. Then we have $r_s=r_{\rm MW}/c_{\rm
MW}(2-\gamma)$ where $r_{\rm MW}\approx 260$ kpc is the virial
radius of MW halo. Then $\rho_s$ can be derived by requiring
$M_{\rm MW}=\int \rho {\rm d}V$. The local density $\rho_\odot$
in this process is checked to be within $0.27$ to $0.25$ GeV cm$^{-3}$
for $\gamma$ varying from $0$ to $1.5$.

Finally there are propagation effects of these particles before being
detected by CR detectors. For charged particles such as positrons and 
antiprotons, we can replace the source term in Eq. (\ref{prop}) with
Eqs. (\ref{annisource})(\ref{decaysource}) and solve the propagation
equations to get the propagated fluxes. For neutral particles like
$\gamma$-ray photons and neutrinos we just need to integrate the 
contribution along the line-of-sight of given direction. More details
of the treatment of $\gamma$-rays can refer to Paper I. Note that 
for neutrinos there will be oscillations between different
flavors. We actually count all flavors of neutrinos from the output
of PYTHIA, and multiply a factor $1/3$ to give the result of muon 
neutrinos.

\subsection{Solar modulation}

The charged particles will interact with the solar wind when entering
the solar system, namely the solar modulation\cite{Itoh}. The force field 
approximation, proposed by Gleeson and Axford \cite{gleeson68}, gives
fairly good description of the solar modulation effects of CRs.
In the force field model charged particle is regarded as entering
an electric field and will lose part of its energy which is generally
described by a potential $\Phi$. For high energy particles ($E_k\gtrsim
20$ GeV) the effect of solar modulation is very weak. For low energy
particles the solar modulation will be important. In our study the 
data of electron spectra from Fermi-LAT and HESS both have energies 
higher than $20$ GeV, and the solar modulation can be safely neglected. 
For the ratio of $e^+/(e^++e^-)$ and $\bar{p}/p$ there are several 
low energy points which may be affected by the solar modulation.
However, the ratio of different types of particles is less sensitive 
to the solar modulation than the flux of single species. In our treatment
only the data of fraction with $E_k\gtrsim 5$ GeV are used, which are
hardly affected by the solar modulation. 

\section{MCMC study of DM parameters}

The full parameter space of this MCMC study is
\begin{equation}
\label{parameter} 
{\bf P} \equiv (m_{\chi}, \langle\sigma v\rangle{\ \rm or\ \tau},
B_{e}, B_{\mu}, B_{\tau}, B_q, \gamma, a_\gamma, b_\gamma, a_{e^-},
b_{e^-}, c_{e^+}, c_{\bar{p}}).
\end{equation}
For the branching ratios we have a normalization condition 
$B_{e}+B_{\mu}+B_{\tau}+B_q\equiv 1$. As a consequence we have $12$
free parameters in total.

\begin{figure}[!htb]
\begin{center}\vspace{10mm}
\includegraphics[width=0.45\columnwidth]{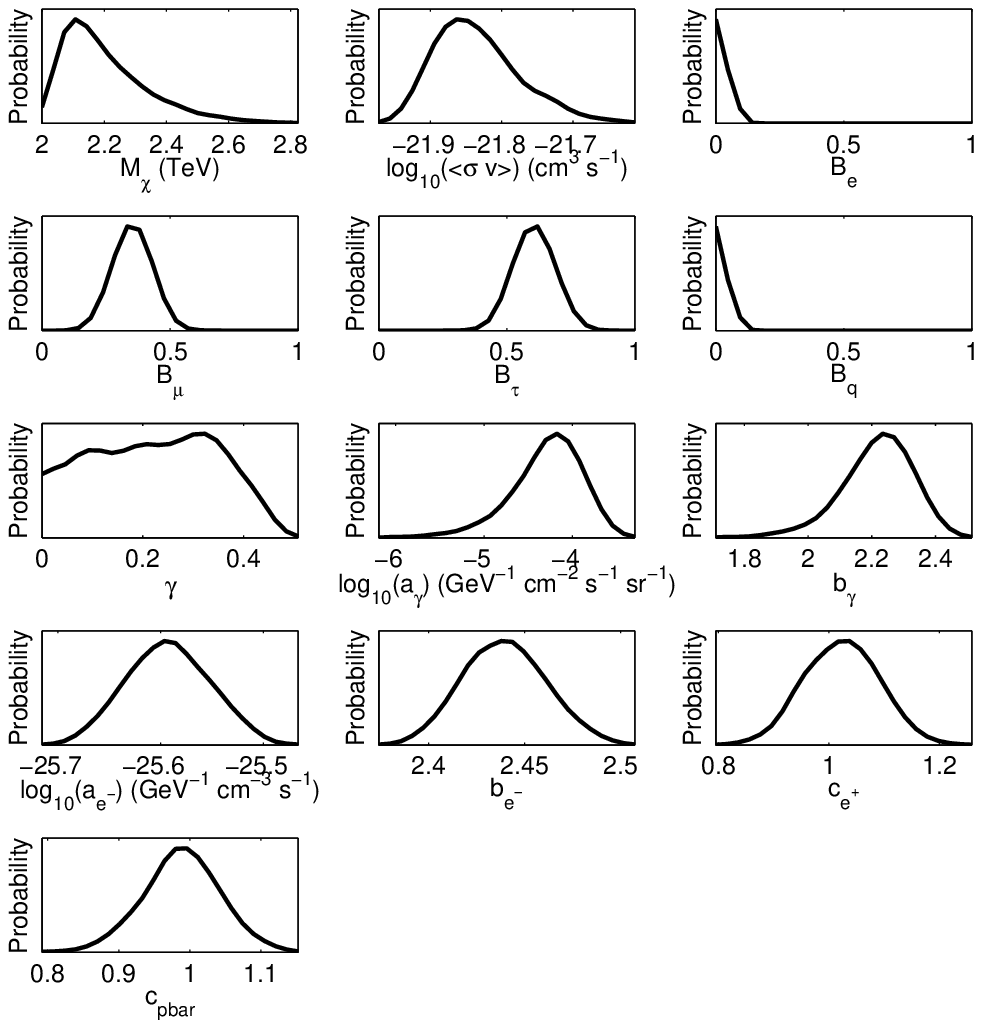}
\hspace{6mm}
\includegraphics[width=0.45\columnwidth]{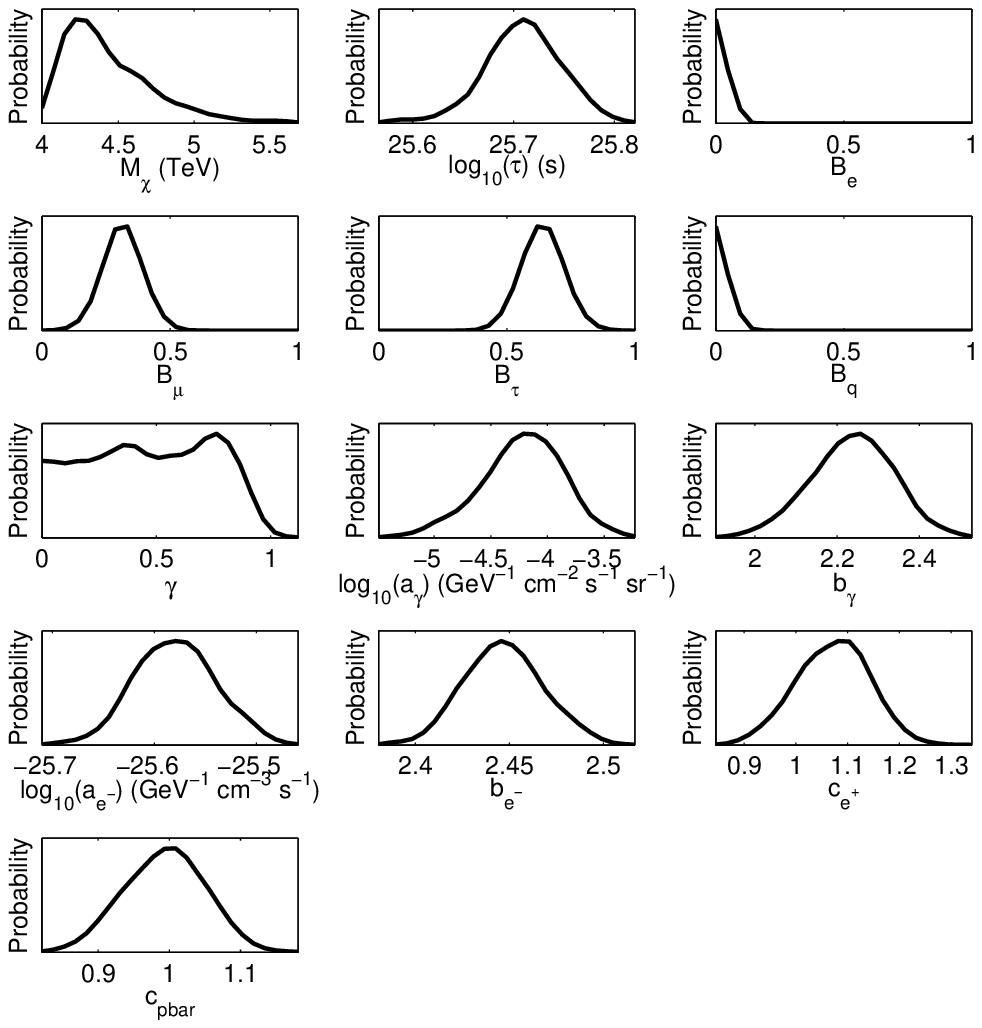}
\caption{Probability distributions of the model parameters in annihilation
({\it left}) and decaying ({\it right}) DM scenarios respectively.
\label{fig:para}}
\end{center}
\end{figure}

The probability distribution of the fitting parameters is shown in
Fig. \ref{fig:para}. The basic fitting results are similar with our
previous study in Paper I. What's new is the constraint on
quark branching ratio from PAMELA $\bar{p}/p$ data. We find that the
$2\sigma$ upper limits of $B_q$ is $0.032$ for DM annihilation and
$0.044$ for DM decay respectively. These results show that the DM to 
explain the recent CR data indeed needs to dominantly couple with 
leptons instead of hadrons. 

As a check of systematics, we calculate the cases of ``MIN'' and ``MAX'' 
propagation parameters. It is found that the basic results do
not differ much from the ``MED'' propagation parameters. The changes
of fitting parameters are generally with several tens percent. The
most sensitive parameter is $c_{e^+}$, which varies from $0.5$ (``MIN'')
to $1.8$ (``MAX''). This change is also expected according to the
results of Ref. \cite{Delahaye:2008ua}. Especially for the quark
branching ratio it is less than $10\%$ in any case.

\section{Neutrino emission}

In this section we discuss the neutrino emission of the DM models.
For each parameter set in the MCMC samples obtained in the previous
calculation, we compute the neutrino flux as a function of energy.
The probability distribution of neutrino flux then can be derived
for each energy bin.

\begin{figure}[!htb]
\begin{center}\vspace{10mm}
\includegraphics[width=0.45\columnwidth]{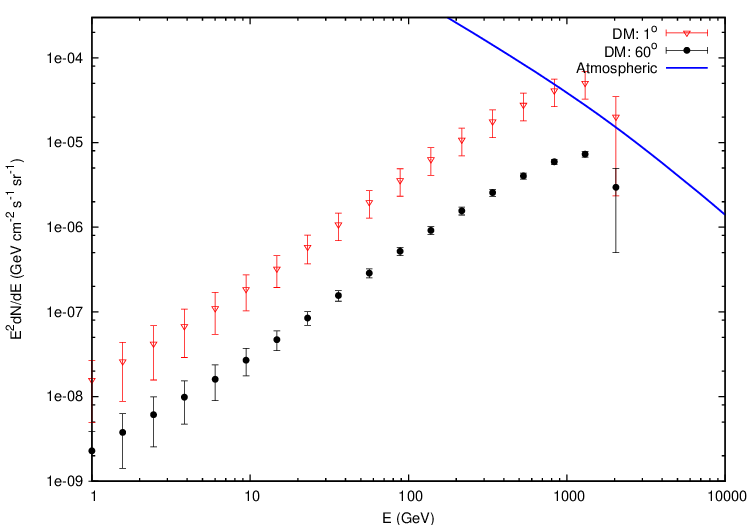}
\includegraphics[width=0.45\columnwidth]{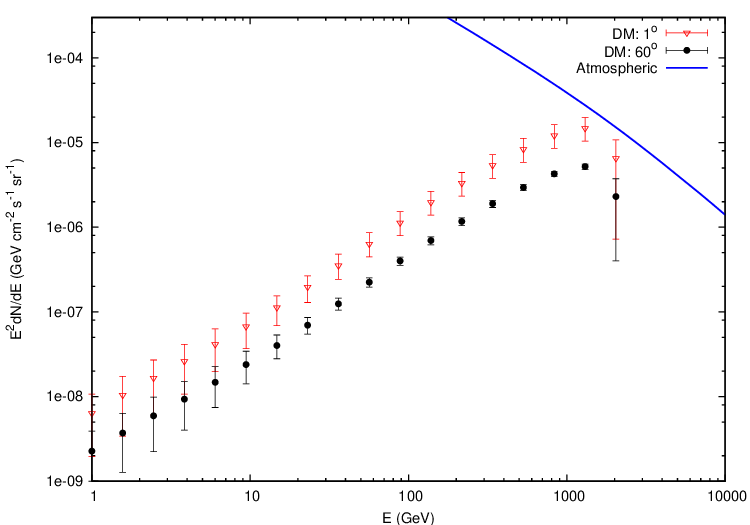}
\caption{Predicted fluxes and $1\sigma$ uncertainties of muon and 
anti-muon neutrinos from the GC for annihilation ({\it left}) and 
decaying ({\it right}) DM scenarios respectively. The blue solid line
shows the atmospheric background.
\label{fig:nuflux}}
\end{center}
\end{figure}

The predicted fluxes of $\nu_{\mu}+\bar{\nu}_{\mu}$ from the GC direction
are shown in Fig. \ref{fig:nuflux}. Two sky regions, $1^{\circ}$ and
$60^{\circ}$ (half angle of the cone) around the GC are calculated to 
show the effects of angular resolution. For comparison we also show
the atmospheric background adopted from Ref. \cite{Honda:2006qj}. The
average results of atmospheric neutrino fluxes are almost the same for
$1^{\circ}$ and $60^{\circ}$ cones.

It is shown that generally the neutrino fluxes are dominated by the
atmospheric background. For the decaying DM scenario the results are
almost always lower than the atmospheric background, even for the
angular resolution as good as $1^{\circ}$, which corresponds to the
best performance of IceCube \cite{Ahrens:2003ix}. For the annihilation
DM case, the neutrino flux from DM will have a chance to exceed the
atmospheric background for energies $\sim$TeV, given a very good 
angular resolution. Otherwise the signal will be still dominated 
by the background.

However, to better see the detectability of such DM-induced neutrinos, 
we need to compare the muon events induced by the neutrinos on a detector.
Following the method given in Ref. \cite{Kistler:2006hp} we calculate
the expect number of muon events on IceCube. Both the contained and 
through-going events are included. The size of IceCube detector is
adopted as $\sim 1$ km$^3$, and the data accumulative time is taken
as $3$ years. The expected muon events from both the atmospheric and 
DM-induced muon neutrinos are compiled in Table
\ref{table:muon}. In the calculation we use the mean values of neutrino
fluxes shown in Fig. \ref{fig:nuflux}. The integral energy for muons is 
adopted from $0.5$ to $2$ TeV, which ensures us to include DM-induced 
signal and exclude atmospheric background as effective as possible. It is 
shown that for a circle sky region with radius $1^{\circ}$ the statistical
significance of DM signal is less than $5\sigma$ for IceCube running
for $3$ years. The case of DM annihilation is some better than DM
decay. If we enlarge the sky region to about $60^{\circ}\,(\pi\ 
{\rm sr})$, the number of events will be much larger, both for 
atmospheric background and DM signal. The significance of DM signals 
can reach $\sim 25-35\sigma$. However, we should keep in mind that in 
this case it would be still difficult to pick out signal events from 
the background events. The information of spatial distribution and 
energy spectroscopy is necessary.

\begin{table}[htb]
\centering
\caption{Number of muon events between $0.5$ and $2$ TeV induced by 
$\nu_{\mu}+\bar{\nu}_{\mu}$ from DM and atmospheric background, for 
$3$-year data taking of IceCube. The number in parentheses is the 
detection significance defined as $N_{\rm DM}/\sqrt{N_{\rm atm}}$.}
\begin{tabular}{cccc}
\hline \hline
cone angle & atm & DM-ann($\sigma$) & DM-decay($\sigma$) \\
\hline
$1^{\circ}$ & $23.4$ & $20.4(4.2)$ & $6.1(1.3)$ \\
$60^{\circ}$ & $7.68\times 10^4$ & $9.71\times 10^3(35.0)$ & $7.07\times 10^3(25.5)$ \\
  \hline
  \hline
\end{tabular}
\label{table:muon}
\end{table}

Compared with the results given in Ref. \cite{Liu:2008ci}, our predicted
neutrino flux for DM annihilation is several times smaller\footnote{Note
this is a rough comparison because the model parameters in Ref. 
\cite{Liu:2008ci} are different from our global fitting ones.}. This is
because in our work the $\gamma$-ray constraint is taken into account
and a smaller value of the slope of DM central cusp $\gamma$ is required. 
Whereas in Ref. \cite{Liu:2008ci} NFW profile is adopted. For decaying
DM scenario the results do not differ much from each other. Our prediction 
of neutrino fluxes should be more self-consistent and more reliable.


\section{Summary}
\label{Sum}

Based on the MCMC code developed in Paper I, we further study the 
properties of DM models which may connect with the recent CR lepton
excesses, after incorporating the hadronic branching ratio and the
constraint from PAMELA $\bar{p}/p$ data. Our global fitting results
show that DM with annihilation or decay channels dominantly to the 
combination of $\tau^+\tau^-$ and $\mu^+\mu^-$ can well describe
the electron and/or positron data measured by PAMELA, Fermi-LAT
and HESS. The mass of DM is about $2$ TeV for annihilation (or $4$ 
TeV for decay) is also consistent with other works taking into account 
the difference of final states \cite{Bergstrom:2009fa,Meade:2009iu}. 
The $\bar{p}/p$ data from PAMELA limit the branching ratio
to quark pairs to be less than $5\%$. This conclusion is qualitatively
consistent with the previous studies \cite{Cirelli:2008pk,Yin:2008bs,
Donato:2008jk}, but should be regarded as the first quantitative
result based on the global fitting method. 

We then calculate the neutrino fluxes from the DM annihilation or decay
according to the MCMC fitting parameters. Due to the  $\gamma$-ray 
constraint on the DM profile, Our expected neutrino flux
for DM annihilation scenario is smaller than that in the previous works 
where the NFW profile is priorly adopted.  
However, as we have shown in Ref. \cite{Bi:2009am} and Paper I, the NFW
profile will be strongly constrained by the diffuse $\gamma$-rays from
the GC ridge observed by HESS \cite{Aharonian:2006au} for DM annihilation
scenario. The derived $2\sigma$ upper limit of the slope of DM central
profile is $\sim 0.5$ in the MCMC study, which is much smaller than the
NFW profile with $\gamma=1$. Although smaller, the neutrino flux 
still could be detected by the forth-coming detector such as IceCube. 
However, a careful analysis about the angular distribution and spectral 
distribution is necessary to separate the DM-induced signal from the 
high atmospheric background. For DM decay scenario the constraint 
from $\gamma$-rays is weaker, and the differences of the neutrino 
flux between this work and other works are smaller. 

\appendix*

\section{Solutions of the propagation equations of positrons and 
antiprotons}

\subsection{Positrons}

The dominant process in the propagation of positrons is energy loss due 
to synchrotron and inverse Compton scattering for energies higher than 
$\sim$GeV. In this paper we will neglect the convection and reacceleration 
of positrons, which is shown to be of little effect for $E\gtrsim 10$ GeV 
(also the interested energy range here) \cite{Delahaye:2008ua}. Then the 
propagation equation is
\begin{equation}
-D\Delta N + \frac{\partial}{\partial E}\left(\frac{{\rm d}E}{{\rm d}t}
N\right)=q({\bf x},E),
\end{equation}
in which the second term in the left hand side represents the energy
losses. The energy loss rate of positrons due to synchrotron and inverse
Compton scattering in the MW can be adopted as ${\rm d}E/{\rm d}t=
-\epsilon^2/\tau_{\rm E}$, with $\epsilon=E/1{\rm\ GeV}$ and $\tau_{\rm E}
\approx 10^{16}$ s \cite{Baltz:1998xv}. We directly write down the
propagator for a point source located at $(r,z)$ from the solar location
with monochromatic injection energy
$E_{\rm S}$ \cite{Lavalle:2006vb,Lavalle:1900wn}
\begin{equation}
{\cal G}_{\odot}^{e^+}(r,z,E\leftarrow E_{\rm S})=\frac{\tau_{\rm E}}
{E\epsilon}\times \hat{\cal G}_{\odot}(r,z,\hat{\tau}),
\label{propposi}
\end{equation}
in which we define a pseudo time $\hat{\tau}$ as
\begin{equation}
\hat{\tau}=\tau_{\rm E}\frac{\epsilon^{\delta-1}-
\epsilon_{\rm S}^{\delta-1}}{1-\delta}.
\end{equation}
$\hat{\cal G}_{\odot}(r,z,\hat{\tau})$ is the Green's function for
the re-arranged diffusion equation with respect to the pseudo time
$\hat{\tau}$
\begin{equation}
\hat{\cal G}_{\odot}(r,z,\hat{\tau})=\frac{\theta(\hat{\tau})}
{4\pi D_0\hat{\tau}}\exp\left(-\frac{r^2}{4D_0\hat{\tau}}\right)
\times {\cal G}^{\rm 1D}(z,\hat{\tau}).
\end{equation}
The effect of boundaries along $z=\pm L$ appears in ${\cal G}^{\rm 1D}$
only. Following Ref. \cite{Lavalle:2006vb} we use two distinct regimes to
approach ${\cal G}^{\rm 1D}$:
\begin{itemize}

\item for $\zeta\equiv L^2/4D_0\hat{\tau}\gg 1$ (the extension of electron
sphere $\lambda\equiv\sqrt{4D_0\hat{\tau}}$ is small)
\begin{equation}
{\cal G}^{\rm 1D}(z,\hat{\tau})=\sum_{n=-\infty}^{\infty}(-1)^n
\frac{\theta(\hat{\tau})}{\sqrt{4\pi D_0\hat{\tau}}}\exp\left(
-\frac{z_n^2}{4D_0 \hat{\tau}}\right),
\end{equation}
where $z_n=2Ln+(-1)^nz$;

\item otherwise
\begin{equation}
{\cal G}^{\rm 1D}(z,\hat{\tau})=\frac{1}{L}\sum_{n=1}^{\infty}
\left[\exp(-D_0k_n^2\hat{\tau})\phi_n(0)\phi_n(z)+
\exp(-D_0k_n'^2\hat{\tau})\phi_n'(0)\phi_n'(z)\right],
\end{equation}
where
\begin{eqnarray}
\phi_n(z)&=&\sin[k_n(L-|z|)];\ \ k_n=(n-1/2)\pi/L,\\
\phi_n'(z)&=&\sin[k_n'(L-z)];\ \ k_n'=n\pi/L.
\end{eqnarray}

\end{itemize}
For any source function $q(r,z,\theta;E_{\rm S})$ the local observed
flux of positrons can be written as
\begin{equation}
\Phi_{\odot}^{e^+}=\frac{v}{4\pi}\times 2\int_0^L{\rm d}z
\int_0^{R_{\rm max}}r{\rm d}r\int_E^{\infty}{\rm d}E_{\rm S}
{\cal G}^{e^+}_{\odot}(r,z,E\leftarrow E_{\rm S})
\int_0^{2\pi}{\rm d}\theta q(r,z,\theta;E_{\rm S}).
\label{fluxposi}
\end{equation}

\subsection{Antiprotons}

It has been shown that for the propagation of antiprotons neglecting
the continuous energy losses and reacceleration can provide a good
enough approach, especially for energies higher than several GeV
\cite{Maurin:2006hy}. We will also adopt this approximation here.
Therefore the relevant processes include the diffusion, convection
and the catastrophic losses --- inelastic scattering and annihilation
in interactions. The propagation equation is
\begin{equation}
-D\Delta N+V_{\rm c}\frac{\partial N}{\partial z}+2h\Gamma_{\rm tot}
\delta(z)N=q({\bf x},E),
\end{equation}
where $\Gamma_{\rm tot}=\sum_{i=H,He}n_i\,\sigma_i^{\bar{p}}\,v$ is
the destruction rate of antiprotons in the thin gas disk with half
height $h\approx 0.1$ kpc \cite{Maurin:2006hy}, $q({\bf x},E)$ is the
source function. The propagator for a point source located at
${\bf x}_{\rm S}$, expressed in cylindrical coordinates $(r,z)$ (symmetric
in $\theta$) is \cite{Maurin:2006hy}
\begin{equation}
{\cal G}^{\bar{p}}_{\odot}(r,z,E)=\frac{\exp(-k_vz)}{2\pi DL}\times\sum_{n=0}
^{\infty}c_n^{-1}K_0\left(r\sqrt{k_n^2+k_v^2}\right)\sin(k_nL)
\sin[k_n(L-z)],
\label{proppbar}
\end{equation}
where $r$ and $z$ are the radial distance and vertical height of the
source, $K_0(x)$ is the modified Bessel function of the second type,
$k_v=V_c/2D$, and $k_n$ is the solution of the equation
$2k_n\cos(k_nL)=-(2h\Gamma_{\rm tot}/D+2k_v)\sin(k_nL)$, and
$c_n=1-\frac{\sin(k_nL)\cos(k_nL)}{k_nL}$. For any source function
$q(r,z,\theta;E)$, the local observed flux is
\begin{equation}
\Phi_{\odot}^{\bar{p}}(E)=\frac{v}{4\pi}\times 2\int_0^L{\rm d}z
\int_0^{R_{\rm max}}r{\rm d}r{\cal G}^{\bar{p}}_{\odot}(r,z,E)
\int_0^{2\pi}{\rm d}\theta q(r,z,\theta;E).
\label{fluxpbar}
\end{equation}


\section*{Acknowledgements}

The MCMC code is adapted from the released COSMOMC code \cite{Lewis:2002ah}. 
We thank Yi-Fu Cai for helpful discussions.
This work is supported in part by National Natural Science Foundation 
of China under Grant Nos. 90303004, 10533010, 10675136 and 10803001 and 
by the Chinese Academy of Science under Grant No. KJCX3-SYW-N2.


\end{document}